\documentstyle[aps,pra]{revtex}

\begin{document}
\title{Implementing universal multi-qubit quantum logic gates in three and
four-spin systems at room temperature}
\date{Aug. 24, 2000}
\author{J. F. Du$^{1,2\thanks{%
djf@ustc.edu.cn}},$ M. J. Shi$^{1,2},$ J. H. Wu$^3$, X. Y. Zhou$^{1,2}$ and
R. D. Han$^{1,2}$}
\address{$^1$Laboratory of Quantum Communication and Quantum Computation, \\
University of Science and Technology of China, Hefei, 230026,\\
P. R. China.\\
$^2$Department of Modern Physics,\\
University of Science and Technology of China, Hefei, 230026,\\
P. R. China.\\
$^3$Laboratory of Structure Biology,\\
University of Science and Technology of China, Hefei, 230026,\\
P. R. China.}
\maketitle

\begin{abstract}
In this paper, we present the experimental realization of multi-qubit gates $%
\Lambda _n\left( not\right) $ in macroscopic ensemble of three-qubit and
four-qubit molecules. Instead of depending heavily on the two-bit universal
gate, which served as the basic quantum operation in quantum computing, we
use pulses of well-defined frequency and length that simultaneously apply to
all qubits in a quantum register. It appears that this method is
experimentally convenient when this procedure is extended to more qubits on
some quantum computation, and it can also be used in other physical systems.

{\bf PACS:03.67.Lx,03.67.-a,76.90.+d}
\end{abstract}

\section{Introduction}

A quantum computer, exploiting quantum state superposition and entanglement,
is capable of performing select computations more powerfully than any
classical computer\cite{divin,chuang4}. Many quantum algorithms that achieve
the same computational tasks much faster than their classical counterparts
have been designed\cite{deutsch,shor,grover}. These theoretical results have
led many groups to try to realize a quantum computer experimentally. Up to
now, there are a rapidly growing number of proposed device technologies for
quantum computation\cite{nmr1,ion,kane,dots}. Of these technologies, those
exploiting nuclear magnetic resonance(NMR) have been the first to
demonstrate non-trivial quantum algorithms with small numbers of qubits\cite
{algorithm1,algorithm2,algorithm3}, most authors realized three-qubit NMR
quantum computation\cite{3bit1,3bit2,3bit3,linden1,kumar}.

One of the key challenges we are facing is to increase the size of the
system used. It is notable that the largest number of qubits used up to date
is seven-qubit\cite{7-bit}. To experimenter, any computational task could be
viewed as a quantum circuit composed of quantum logic gates\cite{Deutsch89},
in analogy to the situation of classical computers. In addition, from the
point of view of decoherence, the circuit must be high efficient in
experiment. Therefore, it is important to demonstrate the Multi-qubit
quantum logic gates on a large number of qubits effectively in practice. On
the other hand, some methods to construct the arbitrary multi-qubit quantum
circuit with a sequence of universal two-qubit gates were reported recently%
\cite{Bar95,cs,group}.

In this paper, we start from the introduction of $\Lambda _n\left(
not\right) $ gate, and try to realize it with a sequence of two-qubit $%
\Lambda _1\left( not\right) $ gates presented by Barenco et al\cite{Bar95}.
Through this procedure we find that in NMR doing nothing (suspending
evolution) in one part of a system while doing something in another part of
the system is nontrivial, i.e. the extension of two-qubit gate to more than
two coupled spins is complicated and not easy to implement. Then, we report
the implementation of $\Lambda _2\left( not\right) $ and $\Lambda _3\left(
not\right) $ gates by utilizing the electromagnetic pulses of well-defined
frequency and length in three and four-qubit systems. For the sake of
experimental convenience, these pulses are physically reasonable choice for
the method of quantum circuit construction in quantum computation rather
than two-qubit $\Lambda _1\left( not\right) $ gate. It seems that this
method is experimentally convenient when we extend this procedure to more
qubits, and one can perform a large class of multi-qubit controlled
rotations simultaneously in many physical systems such as quantum dots,
nuclear spins and trapped ions {\it etc}.

\section{Quantum $\Lambda _n\left( not\right) $ gate}

The $\Lambda _n\left( not\right) $ gate is a $\left( n+1\right) $-qubit
quantum logic gate in which there are $n$ control qubits and one target
qubit, performing the unitary transformation

\begin{equation}
\Lambda _n\left( not\right) \left( \left| x_1,...,x_n,y\right\rangle \right)
=\left| x_1,...,x_n,\bigwedge_{k=1}^nx_k\oplus y\right\rangle
\end{equation}

\[
=\left\{ 
\begin{array}{l}
\left| x_1,...,x_n,y\right\rangle \text{ }\qquad \text{if}%
\bigwedge_{k=1}^nx_k=0 \\ 
\left| x_1,...,x_n,\overline{y}\right\rangle \qquad \text{ if}%
\bigwedge_{k=1}^nx_k=1
\end{array}
\right. 
\]

for all $x_1,...,x_n,y\in \left\{ 0,1\right\} ,$ and $\bigwedge_{k=1}^nx_k$
denotes the AND of the boolean variables $\left\{ x_k\right\} .$ It can be
seen that $n$ control qubits does not change their values $\left|
x_1,...,x_n\right\rangle $ after the action of the $\Lambda _n\left(
not\right) $ gate and the target qubit flips its state $\left|
y\right\rangle $ to $\left| \overline{y}\right\rangle $ if and only if $%
\bigwedge_{k=1}^nx_k=1$. The $\Lambda _n\left( not\right) $ gate can also be
described by the following unitary matrix

\begin{equation}
\Lambda _n\left( not\right) =\left( 
\begin{array}{ll}
I_{2^{n+1}-2} & 0 \\ 
0 & \sigma _x
\end{array}
\right)
\end{equation}

where the basis states of $n+1$ qubits are lexicographically ordered, i.e., $%
\left| 0\cdots 00\right\rangle ,\left| 0\cdots 01\right\rangle ,$ $\cdot
\cdot \cdot \left| 1\cdots 11\right\rangle $, $I_{2^{n+1}-2}$ is the $\left(
2^{n+1}-2\right) \times \left( 2^{n+1}-2\right) $ identity matrix and $%
\sigma _x$ are Pauli spin matrices.

when $n=1$, $\Lambda _1\left( not\right) $ is the so-called CNOT gate

\begin{equation}
\Lambda _1\left( not\right) =\left| 00\right\rangle \left\langle 00\right|
+\left| 01\right\rangle \left\langle 01\right| +\left| 10\right\rangle
\left\langle 11\right| +\left| 11\right\rangle \left\langle 10\right|
=\left( 
\begin{array}{ll}
I_2 & 0 \\ 
0 & \sigma _x
\end{array}
\right)
\end{equation}

It has a role to play in quantum measurement, in creation and manipulation
of entanglement and in quantum error correction. Furthermore, $\Lambda
_1\left( not\right) $ gate is of central importance in quantum computation
because any quantum logic gates can be decomposed into a sequence of
one-qubit rotations and two-qubit $\Lambda _1\left( not\right) $ gates\cite
{Bar95}. Most implementation of the $\Lambda _1\left( not\right) $ gate in a
liquid state NMR involved a special sequence of resonance electromagnetic
pulses. For example, in two-spin physical system of the carbon-13 labeled
chloroform, the ``modified'' $\Lambda _1\left( not\right) $ gate was
implemented using two radio-frequency pulses\cite{chuang2}. The first pulse
induced a $\pi $/$2$ rotation of the target spin around the one axis of the
rotating reference frame. The second pulse induced a similar rotation around
the other axis. The delay time between two pulses was $\pi /2J$, where $J$
is the interaction constant. This action is modified in the sense that it
differs from $\Lambda _1\left( not\right) $ gate in relative phases but it
require even less resources. One can also see the ideal implementation of $%
\Lambda _1\left( not\right) $ gate with more resources\cite{chuang,du}.

If $n=2$, $\Lambda _2\left( not\right) $ can be regarded as a three-qubit
quantum logic gate which is named as Toffolli-gate now, and this gate plays
the role of a universal gate for reversible circuits.

\begin{equation}
\Lambda _2\left( not\right) =\stackrel{101}{\sum_{x=000}}\left|
x\right\rangle \left\langle x\right| +\left| 110\right\rangle \left\langle
111\right| +\left| 111\right\rangle \left\langle 110\right| =\left( 
\begin{array}{ll}
I_6 & 0 \\ 
0 & \sigma _x
\end{array}
\right)
\end{equation}

Similarly, the four-qubit gate $\Lambda _3\left( not\right) $ can be written
as

\begin{equation}
\Lambda _3\left( not\right) =\stackrel{1101}{\sum_{x=0000}}\left|
x\right\rangle \left\langle x\right| +\left| 1110\right\rangle \left\langle
1111\right| +\left| 1111\right\rangle \left\langle 1110\right| =\left( 
\begin{array}{ll}
I_{14} & 0 \\ 
0 & \sigma _x
\end{array}
\right)
\end{equation}

One method to implement $\Lambda _2\left( not\right) $ gate and $\Lambda
_3\left( not\right) $ gate would be to make use of one-qubit rotations and
two-qubit $\Lambda _1\left( not\right) $ gates. several schemes have been
proposed to realize $\Lambda _n\left( not\right) $ gates by these
foundational quantum logic gates. However, extending the number of qubits
has not been proved easy. Barenco et al. exhibited to obtain the best known
three-qubit Toffoli gate\cite{toffoli}$\left( \Lambda _2\left( not\right)
\right) $ with five $\Lambda _1\left( not\right) $ gates; in this same
paper, they also built up a $\Lambda _3\left( not\right) $ gate with
thirteen $\Lambda _1\left( not\right) $ gates\cite{Bar95}. Reck {\it et al.}
proved that any $n$-qubit quantum circuit, expressed as an unitary operator,
could be constructed using a finite number $\left( \Theta \left(
n^34^n\right) \right) $ of two-qubit gates\cite{reck94}. More important, due
to the interactions among the qubits, to apply two-qubit gate while doing
nothing on the other $\left( n-2\right) $ qubits in $n$-qubit system is much
more complicated than to apply the same two-qubit gate in a two-qubit NMR
quantum computer\cite{linden}.

The above scheme to realize the $\Lambda _n\left( not\right) $ gates $\left(
n\geq 2\right) $ depends heavily on using the fundamental two-qubit $\Lambda
_1\left( not\right) $ gate. Although each gate in these schemes can be
implemented by appropriately refocused evolutions in NMR, however, the
approach to realize $\Lambda _n\left( not\right) $ gate in the n-qubit NMR
system according to these theoretical schemes is not efficient in general.
There are two main reasons. One is that most of these schemes are to
construct the arbitrary quantum circuit with a sequence of the two-qubit
operation $\Lambda _1\left( not\right) $. This may not be a physically
reasonable choice in some NMR quantum computation, but for the moment this
should be considered as a mathematical convenience which will permit us to
address somewhat general questions. The other is that NMR quantum operations
depend heavily on the molecule being used. The fact that the chemical shift
and coupling constant are small in the spin systems leads to the requirement
of the highly frequency selectivity which is important in quantum operation.

\section{Experimental realization}

Seth Lloyd described a method to perform arbitrary quantum circuit by a
sequence of electromagnetic pulses of well-defined frequency and length in a
weakly-coupled quantum systems\cite{lloyd1,lloyd2}. For the purpose of
interpreting this method, let us consider a macroscopic ensemble of $n$-spin
molecules at room temperature. The $n$ nuclear spins in each molecule
represent a $n$-qubit register. The qubits are labeled by the characteristic
frequencies, $\omega _k$, $\left( k=\text{1 to n}\right) $ due to the Zeeman
interaction of the nuclear spins with the magnetic field. The Hamiltonian of
the nuclear spins in a molecule in solution is well approximated for the
following reason: small interactions with other nuclei do not play a major
role in the dynamics while higher order terms in the spin-spin coupling can
be disregarded in the first-order model, and the rapid molecular tumbling
averages away in the liquid at a high magnetic field.

\begin{equation}
H_{n\ qubits}=-\frac \hbar 2\left[ \stackrel{n}{\sum_{k=1}}\omega _k\sigma
_z^k+\pi \sum\limits_{k>m}J_{m,k}\sigma _z^m\sigma _z^k\right]
\end{equation}

where $\omega _k$ is the Lamour frequency of spin $k$ and $\sigma _z^k$ is
the $\widehat{z}$ Pauli operator of spin $k$. $J_{m,k}$ is the strength of
scalar weakly coupling between $k$ spin and $m$ spin. Because the energy
level differences are small at room temperature $E_k/K_BT\ll 1$ and scalar
coupling is weak $J\ll \left| \omega _{k+1}-\omega _k\right| $; the initial
state is the thermal equilibrium state,

\begin{equation}
\rho _{eq}=\frac{\varrho ^{-E_k/K_BT}}{\sum_{k=0}^{2^n+1}\varrho ^{-E_k/K_BT}%
}
\end{equation}
we use the complete set of the basis states, $\{\left| 0_1\cdots
0_{n-1}0_n\right\rangle $ $=\left| 0\right\rangle ,$ $\left| 0_1\cdots
0_{n-1}1_n\right\rangle $ $=\left| 1\right\rangle ,$ $\cdot \cdot \cdot ,$ $%
\left| 1_1\cdots 1_{n-1}1_n\right\rangle $ $=$ $\left| 2^n-1\right\rangle \}$%
, form a complete set of eigenstates of Hamiltonian $H$: $H\left|
k\right\rangle $ $=E_k\left| k\right\rangle ,$ $\left( k=0,\cdot \cdot \cdot
,2^n-1\right) $. For example, the energies of the ground state, $\left|
0\cdots 00\right\rangle $, and the excited state, $\left| 1\cdots
11\right\rangle $, are 
\begin{equation}
E_0=-\frac \hbar 2\stackrel{n}{\sum_{k=1}}\left( \omega _k+\pi
\sum\limits_{k>m}J_{m,k}\right)
\end{equation}

\[
E_{2^n-1}=-\frac \hbar 2\stackrel{n}{\sum_{\alpha =1}}\left( -\omega _k+\pi
\sum\limits_{k>m}J_{m,k}\right) 
\]

we assume that: $\omega _1<\omega _2<\omega _3,$ $\cdot \cdot \cdot ,<\omega
_n,\ $and $J\ll $ $\omega _{k+1}-\omega _k,$ ($k=1,2,$ $\cdot \cdot \cdot ,$ 
$2^n-1$).

Assume that the frequency of the external magnetic field, $\omega $, is
resonant with the frequency of the transition, $\left| 1_1\cdots
1_{n-1}0_n\right\rangle $ $\leftrightarrow $ $\left| 1_1\cdots
1_{n-1}1_n\right\rangle $, and that this transition frequency is different
from all other single-spin transition frequencies. In this case, by applying
a single $\pi $-pulse with frequency $\omega $, we could implement the
modified quantum $\Lambda _{n-1}\left( not\right) $ gate in an ensemble of $%
n $-spin molecules,

\begin{equation}
U=\varrho ^{-i\pi \left[ \frac 1{16}\left( 1-\sigma _z^1\right) ,\cdot \cdot
\cdot ,\left( 1-\sigma _z^{n-2}\right) \left( 1-\sigma _z^{n-1}\right)
\sigma _x^n\right] }=\left( 
\begin{array}{ll}
I_{2^n-2} & 0 \\ 
0 & -i\sigma _x
\end{array}
\right)
\end{equation}

The above operation changes the state of the $n$-th spin (the target spin): $%
\left| 0\right\rangle \leftrightarrow -i\left| 1\right\rangle $, only if the
state of any other spins is $\left| 1\right\rangle $. Some groups tested
this method experimentally in their small two-qubit and three-qubit NMR
quantum computers\cite{linden1,kumar,cory}. Here, we extend this method to
four-qubit systems and realized the Toffoli gate and $\Lambda _3\left(
not\right) $ gate experimentally.

\subsection{three-qubit Toffoli gate}

The single-pulse experimental realization of a modified three-qubit quantum $%
\Lambda _2\left( not\right) $ gate (Toffoli gate) is shown in Figure 1. In
our NMR approach to realize three-qubit $\Lambda _2\left( not\right) $ gate
we need three qubits. A convenient system we used is the liquid consisting
of identical chlorostyrene molecules, the three weakly coupled spin 1/2
hydrogen nuclei, which are distinguished by their different resonance
frequencies, i.e. $\omega _1$, $\omega _2$and $\omega _3$, can be regarded
as the isolated qubits ($\omega _1<\omega _2<\omega _3$).

In this experiment, the difference frequencies of the three spins are $%
\omega _3-\omega _2=471.8Hz$ and $\omega _3-$ $\omega _1=700.3Hz$. The
corresponding $J$ coupling constant between the two qubits are $%
J_{13}=10.9Hz $, $J_{23}=17.6Hz$, and $J_{12}=0.65Hz$. The transmitter
frequency of the transition-pulse is set at the $\omega _3/2\pi
+J_{13}/2+J_{13}/2$. The power and the duration of the transition-pulse are
adjusted to optimize the frequency selectivity. The real part of the Fourier
transform of the resulting signal gives a NMR spectrum of the target qubit-3
containing four peaks at frequencies $\omega _3/2\pi \pm J_{13}/2\pm
J_{13}/2 $. these four peaks (from left to right) correspond to the two
control qubits in the following states $\left| 00\right\rangle ,\ \left|
01\right\rangle ,\ \left| 10\right\rangle \ $and $\left| 11\right\rangle $.
It is clear that our implementation of the quantum $\Lambda _2\left(
not\right) $ gate leaves the computer in a final result as expected: if any
one of the two control qubits is in ground state , the target qubit-3 does
not change its value after the action of the gate; and if the two control
qubits are both in states $\left| 1\right\rangle $ , the target qubit-3
flips its state $\left| 0\right\rangle \leftrightarrow \left| 1\right\rangle 
$.

\subsection{four-qubit $\Lambda _3\left( not\right) $ gate}

The experimental realization of a modified four-qubit quantum $\Lambda
_3\left( not\right) $ gate is shown in Figure 2. The physical system that we
used to implement the $\Lambda _3\left( not\right) $ gate is the
liquid-state proton and carbon NMR of $^{13}C_3$ labeled alanine $NH_3^{+}-$ 
$C^{^\alpha }H\left( C^{^\beta }H_3\right) $ $-$ $C^{\acute{\prime}}OOH$,
with decoupling of the methyl protons, alanine exhibits a weakly coupled
four-spin system. The $C^{^\alpha }$ was chosen as target qubit (labeled
qubit-4) because of its well-resolved couplings with the other carbons $%
\left( J_{C^{^\alpha }C^{\acute{\prime}}}=53.92Hz,\text{ }J_{C^{^\alpha
}C^{^\beta }}=34.42Hz\right) $ and with the adjacent proton $\left(
J_{C^{^\alpha }H}=144.89Hz\right) $, which were chosen as the three control
spins (labeled qubit-1, qubit-2 and qubit-3). The selective $\pi $ pulse
were applied on the connected single-quantum transition $\left|
1_11_21_30_4\right\rangle \leftrightarrow $ $\left|
1_11_21_31_4\right\rangle $ of the target qubit-4, it is noted that all
eight lines arise from the target qubit-4, and the intensity of each line is
proportional to the total population difference between the corresponding
states. It is clear that our implementation of the quantum logic gate leaves
the computer in a final result as expected,

All NMR experiments are performed at room temperature and pressure on Bruker
Avance $DMX500$ spectrometer $\left( 11.7T\right) $ with a $5mm$ probe in
Laboratory of Structure Biology, University of Science and Technology of
China. We can see the small but significant distortions in the final spectra
arising from the difficulty of implementing perfect selective pulses and
inhomogeneity of the magnetic field. The experimental task is to shape the
radio-frequency pulse envelope so as to achieve sufficient selectivity in
the frequency domain that there is negligible perturbation of the next
nearest neighbor of the spin multiplet.

\section{Conclusion}

In conclusion, utilizing the proper resonant pulse of the well-defined
frequency and length, we have demonstrated the reversible and universal
three and four-qubit quantum $\Lambda _n\left( not\right) $ logic gates in
two-level nuclear spin systems at room temperature.

One of the key challenges in quantum computation is to try to increase the
size of the system used. The transition pulse that used in this paper can
apply to all qubits of a quantum register in one operation. the operational
time of the pulse is not influenced by the number of the qubits, therefore,
it is experimentally convenient when we extend this procedure to more
qubits. This advantage may be helpful to answer the question given a unitary
transformation corresponding to a quantum computational task, what is the
shortest sequence of pulses and evolutions which generates it? Further, in
quantum information processing, this single-pulse quantum operation is
useful not only in an ensemble of nuclear spins system but also in many
other physical systems such as quantum dots, trapped ions {\it etc.}

Although the multi-qubit gates implementation with transition pulse is much
simpler, it must be noted that it takes more time than the multi-qubit gates
implementation with spin-selective pulse. For the experimenters, a
reasonable choice should be to combine the utility of various NMR methods.

{\bf ACKNOWLEDGMENTS}

This project was supported by the National Nature Science Foundation of
China.

{\bf Figure Captions:}

Fig. 1, The $\Lambda _3\left( not\right) $ quantum gate implemented on the
four-spin system of alanine. $\left( a\right) $ experimental NMR spectra of
the target qubit of the input thermal equilibrium state. $\left( b\right) $
experimental NMR spectra of the target qubit of the output state of the $%
\Lambda _3\left( not\right) $ gate.

Fig. 2, The $\Lambda _3\left( not\right) $ quantum gate implemented on the
four-spin system of alanine. $\left( a\right) $ experimental NMR spectra of
the target qubit of the input thermal equilibrium state. $\left( b\right) $
experimental NMR spectra of the target qubit of the output state of the $%
\Lambda _3\left( not\right) $ gate.

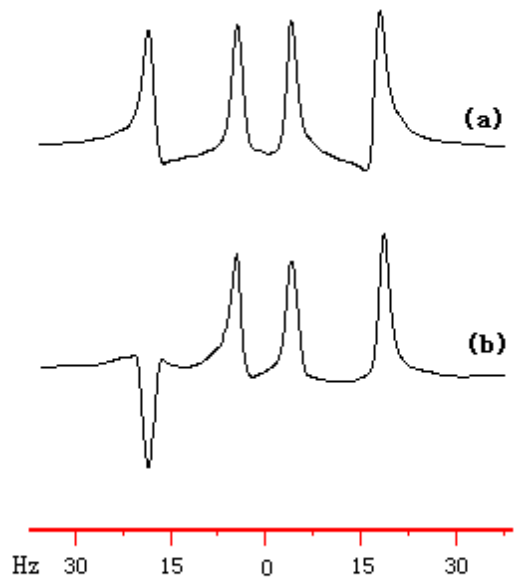
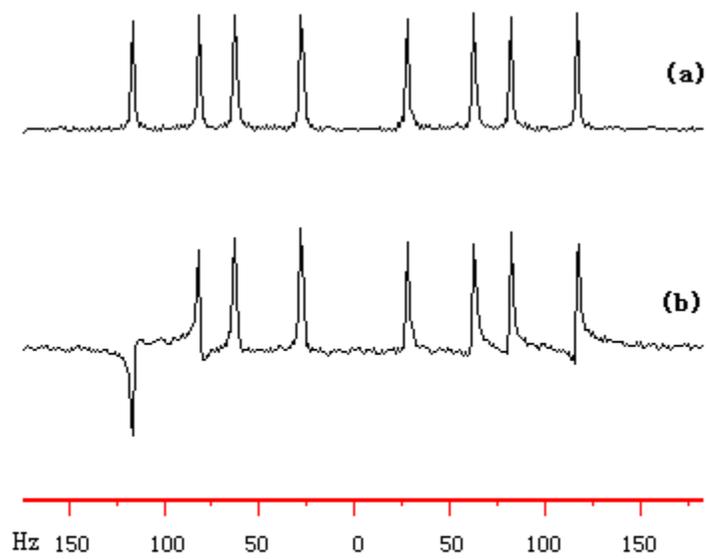
\end{document}